\documentclass[english]{sbrt}
\usepackage[english]{babel}
\usepackage[utf8]{inputenc}
\usepackage{amsmath}
\usepackage{amssymb}
\usepackage{bm}
\usepackage{graphicx}
\usepackage{svg}
\usepackage{tikz}
\usepackage{circuitikz}
\usetikzlibrary{arrows.meta,circuits.ee.IEC,intersections,shapes.geometric}
\usepackage{tkz-euclide}

\usepackage{xcolor}
\definecolor{milacgreen}{HTML}{10ba68}
\definecolor{milacred}{HTML}{ff4d4d}
\definecolor{darkgreen}{HTML}{20684a}
\definecolor{milacblue}{HTML}{7ab2ff}
\definecolor{darkblue}{HTML}{083f8a}
\definecolor{white_soft}{HTML}{FFFFFF}
\definecolor{yellow_light}{HTML}{FFF2CC}
\definecolor{yellow_dark}{HTML}{8A6500}
\definecolor{pink_light}{HTML}{FCE4EC}
\definecolor{pink_dark}{HTML}{8A2042}
\definecolor{blue_light}{HTML}{E3F2FD}
\definecolor{blue_dark}{HTML}{083F8A}
\definecolor{red_light}{HTML}{FFD6D6}
\definecolor{green_dark}{HTML}{20684A}
\definecolor{yellow_mid}{HTML}{a8924f}
\definecolor{pink_mid}{HTML}{C94F7C}
\definecolor{blue_mid}{HTML}{356AA0}
\newcommand{\chg}[1]{#1}

\begin{document}
\raggedbottom

\title{Projection-Based Partitioned MiLAC Architecture for MU-MIMO Precoding}

\author{Israel Coelho Cunha, Francisco L. F. Martins, Giovanni G. R. Milan, Fazal Asim and André L. F. de Almeida 
\thanks{This work is partially supported by the National Institute of Science and Technology (INCT-Signals) sponsored by Brazil’s National Council for Scientific and Technological Development (CNPq) (Proc. 406517/2022-3), and FUNCAP (Proc. INCT-25255-82587.32.41/64).}%
}

\maketitle

\markboth{XLIV SIMPÓSIO BRASILEIRO DE TELECOMUNICA\c{C}\~{O}ES E PROCESSAMENTO DE SINAIS - SBrT 2026, 29 DE SETEMBRO À 2 DE OUTUBRO DE 2026, SALVADOR, BA}{}


\begin{abstract}
This work proposes a partitioned Microwave Linear Analog Computer (MiLAC) architecture for user-specific MU-MIMO precoding based on independently configured analog subnetworks. The proposed method reformulates user-level interference suppression as an invertible projection-based transformation, which is computed from the channel and synthesized into the MiLAC admittances. While the resulting offline configuration incurs higher computational complexity than conventional ZFBF and regularized ZFBF, it remains substantially below fully digital Block Diagonalization. Once configured, the Sub-MiLACs perform the precoding operation in the analog domain, yielding significantly lower online processing complexity. Under ideal conditions, the proposed method achieves performance close to fully digital BD.
\end{abstract}

\begin{keywords}
MU-MIMO, block diagonalization, MiLAC.
\end{keywords}

\section{Introduction}

Multiuser multiple-input multiple-output (MU-MIMO) systems enable simultaneous transmission to multiple users over the same time-frequency resources, significantly improving spectral efficiency in modern wireless communications \cite{tse2005fundamentals,bd}. However, in MU-MIMO systems without inter-user coordination on the receiver side, performance strongly depends on the transmitter's ability to suppress inter-user interference while maintaining manageable computational complexity.

Classical precoding techniques such as block diagonalization (BD) and zero-forcing beamforming (ZFBF) mitigate interference through digital matrix operations involving pseudoinverses or singular value decompositions (SVDs) \cite{bd,yoo2006optimality}. Although these approaches can provide excellent performance, their computational cost increases rapidly with the number of antennas and users, making real-time implementation challenging in large-scale systems \cite{bjornson2025gigantic}. Hybrid analog/digital beamforming has therefore been explored as a lower-complexity alternative for large-antenna and millimeter-wave systems \cite{asim2023hybrid}.

Recently, microwave linear analog computer (MiLAC) architectures have emerged as a promising approach for low-complexity beamforming and precoding in wireless communication systems \cite{nerini2025part1,nerini2025part2,nerini2025capacity}. A MiLAC is typically modeled as a multiport microwave circuit composed of tunable admittance elements that interconnect a set of excited and unexcited ports. By properly configuring these admittances, the electromagnetic response of the network can be shaped to realize desired linear transformations between the input excitations and the resulting port voltages.

This capability enables MiLAC-aided beamforming architectures, in which part of the beamforming and interference-suppression operations are embedded directly into the analog microwave network, thereby reducing the amount of run-time digital processing required for precoding.

A useful interpretation of MiLAC-based beamforming is that the microwave
network can be configured to implement a desired linear transformation
obtained during an offline configuration stage. In this approach, the
channel-dependent transformation is first determined digitally and then
embedded into the tunable admittance parameters of the MiLAC. During
transmission, the configured network applies this transformation directly to
the incoming signals through its electromagnetic response, thereby shifting
the corresponding linear operation from the online digital domain to the
analog hardware.

This distinction is particularly relevant for MU-MIMO precoding. While the
digital computation required to configure a MiLAC remains channel-dependent,
it needs to be performed only when the network configuration is updated.
Once configured, the analog network can execute the corresponding linear
transformation without performing the associated matrix operations for every
transmitted symbol. This separation between offline configuration and online
analog execution provides a means of reducing the digital processing burden
of real-time precoding.

However, directly applying this approach to MU-MIMO precoding remains
challenging when the desired transformation consists of multiple
user-dependent operations. In particular, each user's transmitted
contribution must be constrained with respect to the interference channels of
all remaining users. Classical Block Diagonalization (BD) achieves this by
constructing a user-specific null-space basis, typically through singular
value decomposition (SVD) \cite{bd}. Although such a precoder could in
principle be synthesized into a MiLAC after being computed digitally, a
single global MiLAC transformation does not naturally expose the independent
user-specific structure required by BD.

This motivates the proposed partitioned MiLAC formulation, in which the
global analog network is decomposed into independent user-specific
Sub-MiLAC subnetworks. For each user, a regularized projection operator is
computed during the offline configuration stage and synthesized into the
corresponding Sub-MiLAC. During online operation, the subnetwork directly
applies this projection to the dimension-matched excitation associated with
the user's symbols. The proposed formulation therefore preserves the
user-specific interference-nulling structure of BD while avoiding explicit
null-space basis extraction and SVD-based processing during the online
transmission stage.

The resulting architecture does not eliminate the channel-dependent digital
configuration cost. Instead, it shifts the computational burden associated
with user-specific interference suppression from repeated online digital
processing to an offline configuration stage, after which the projection is
executed by the analog microwave subnetworks.

The main contributions of this work are summarized as follows:
\begin{itemize}
\item A partitioned Sub-MiLAC architecture that exposes independent
user-specific analog processing paths for MU-MIMO precoding.

\item A MiLAC-compatible regularized projection formulation that replaces
explicit null-space basis extraction by directly synthesizable projection
transformations.

\item An offline/online processing framework in which the channel-dependent
projection matrices are computed during configuration and the resulting
interference-suppression transformations are executed by the analog
Sub-MiLAC subnetworks during transmission.
\end{itemize}

\begin{figure}
    \begin{center}
	{\input{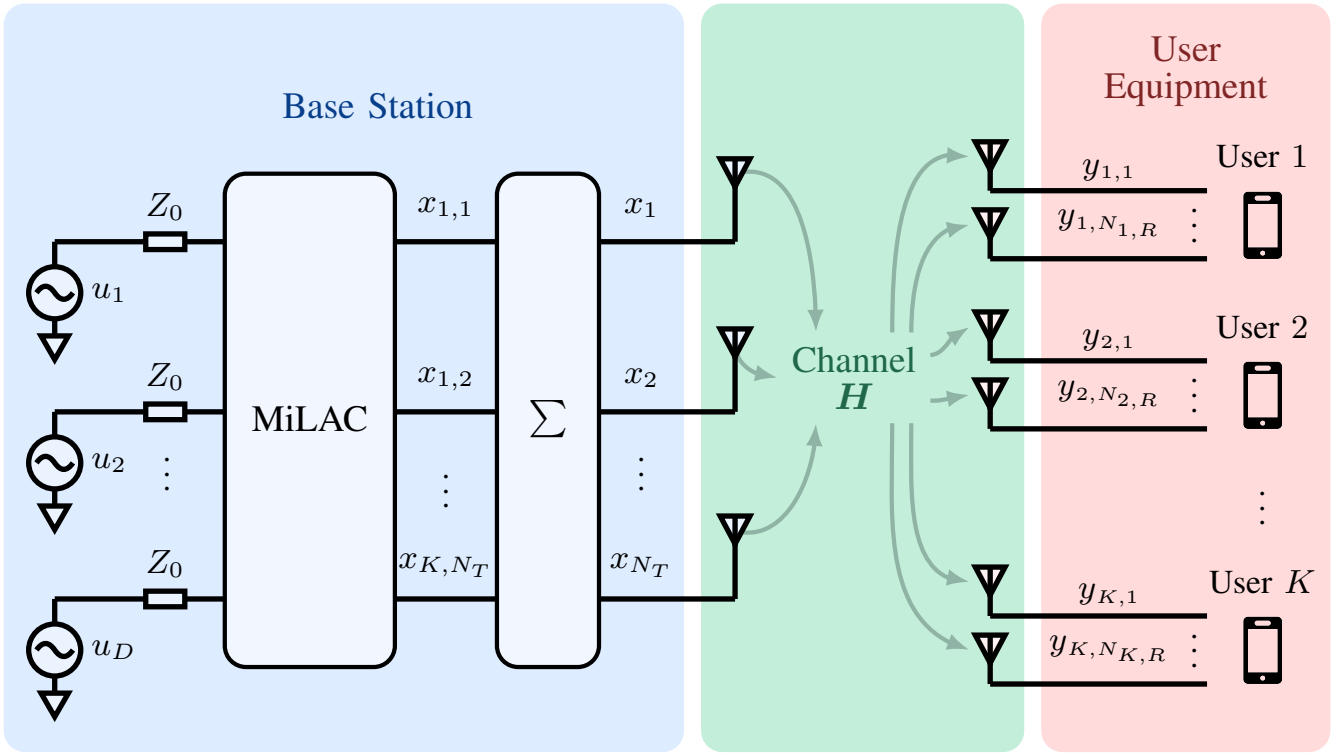}}
    {\footnotesize \caption{\label{fig:system_model}System model of the considered MU-MIMO downlink scenario employing a MiLAC-based analog precoding architecture.}}
    \end{center}
    \vspace{-0.4cm}
    
\end{figure}
\section{System Model}
Consider a wireless flat-fading MU-MIMO downlink scenario in which a single base station (BS), equipped with $N_T$ transmit antennas, simultaneously serves $K$ user equipments (UEs), where the $k$-th UE is equipped with \chg{$N_{R,k}$} receive antenna elements. Prior to transmission through the antenna array, the BS employs a $P$-port MiLAC-designed beamforming precoder.
Fig.~\ref{fig:system_model} illustrates the considered MU-MIMO communication system and the associated MiLAC-based analog precoding architecture.

Let \chg{$\bm{s}_k \in \mathbb{C}^{D_k}$} denote the data-symbol vector intended for user $k$, where \chg{$D_k$} is the number of data streams assigned to that user. The global symbol vector is
{
$
    \bm{s}
    = [\bm{s}_1^T, \bm{s}_2^T, \ldots, \bm{s}_K^T]^T
    \in \mathbb{C}^{D\times1},
    \label{eq:global_symbol_vector}
$
}
with \chg{$D=\sum_{k=1}^{K}D_k$}.

{
Before transmission, each user's symbol vector, $\bm{s}_k$, is mapped from the stream domain to the transmit-antenna domain through the dimension-matching preprocessing detailed in Section~\ref{subsec:dimension_matching}. The resulting vector, $\bm{u}_k \in \mathbb{C}^{N_T\times 1}$, is the MiLAC excitation associated with user $k$.

The full excitation vector is then formed by concatenating the excitation vectors associated with each user and is applied to the directly excited MiLAC ports. It is given by
\begin{equation}
    \bm{u} =
    \left[
        \bm{u}_1^{T},
        \bm{u}_2^{T},
        \ldots,
        \bm{u}_K^{T}
    \right]^{T}
    \in
    \mathbb{C}^{KN_T\times1}.
\end{equation}

By properly configuring the admittance values of the MiLAC network, its electromagnetic response generates the user-specific precoded transmit contributions $\bm{x}_k\in\mathbb{C}^{N_T\times1}$. The physical interpretation of the excitation and voltage-response ports is detailed in Section~III and Fig.~\ref{fig:submilac_architecture}(b).
}

The final transmitted signal\chg{, $\bm{x}$,} applied to the BS antenna array is obtained by superposing all user-specific contributions\chg{, $\bm{x}_k$,} trough the $\sum$ block in Fig.~\ref{fig:system_model}, i.e.,
{
\begin{equation}
    \bm{x}
    =
    \sum_{k=1}^{K}
    \bm{x}_k\in
    \mathbb{C}^{N_T\times1}.
\end{equation}
}

The resulting transmitted signal can be represented as
{
\begin{equation}
    \bm{x}
    =
    \bm{W}\bm{u}
    ,
    \label{eq:transmitted_signal}
\end{equation}
}
where
{
$
\bm{W}
\in
\mathbb{C}^{N_T\times KN_T}
$
}
denotes the effective precoding transformation implemented by the MiLAC network.

The overall precoding matrix is partitioned as
{
$
\bm{W}
=
[\bm{W}_1,\bm{W}_2,\ldots,\bm{W}_K],
$
}
where
{
$
\bm{W}_k
\in
\mathbb{C}^{N_T\times N_T}
$
}
denotes the precoding transformation associated with user $k$.

The transmitted signal then propagates through the wireless channel toward each user. The channel between the BS and user $k$ is modeled by \chg{$\bm{H}_k \in \mathbb{C}^{N_{R,k}\times N_T}$}, which maps the transmitted vector from the BS antenna space into the receive-antenna space of the $k$-th user.

{
The aggregate MU-MIMO channel,
$\bm{H} \in \mathbb{C}^{N_R \times N_T}$,
is formed by vertically concatenating the channel matrices of all users, where
$N_R=\sum_{k=1}^{K} N_{R,k}$ is the total number of receive antennas across all users.
}

The received signal at user $k$ is
{
\begin{equation}
    \bm{y}_k
    = \bm{H}_k\bm{x}+\bm{n}_k
    \in \mathbb{C}^{N_{R,k}\times 1},
    \label{eq:received_signal}
\end{equation}
}
where \chg{$\bm{n}_k\in\mathbb{C}^{N_{R,k}}$} denotes the additive noise vector at user $k$. Substituting (\ref{eq:transmitted_signal}) in (\ref{eq:received_signal}) gives
{
\begin{equation}
    \bm{y}_k
    = \bm{H}_k\bm{W}_k\bm{u}_k
    + \bm{H}_k\tilde{\bm{W}}_k\tilde{\bm{u}}_k
    + \bm{n}_k.
\end{equation}
Here, $\bm{H}_k\bm{W}_k\bm{u}_k$ is the desired received signal, whereas $\bm{H}_k\tilde{\bm{W}}_k\tilde{\bm{u}}_k$ is the aggregate interference produced at user $k$ by the complementary excitation, $\tilde{\bm{u}}_k$, and precoding blocks, $\tilde{\bm{W}}_k$, associated with all other users.
}

{
The complementary channel matrix associated with user $k$ is obtained by
stacking the channel matrices of all remaining users,
$
\tilde{\bm{H}}_k =
\begin{bmatrix}
\bm{H}_1^T,
\cdots,
\bm{H}_{k-1}^T,
\bm{H}_{k+1}^T,
\cdots,
\bm{H}_K^T
\end{bmatrix}^{T}.
\label{eq:aggregated_channel}
$
}
To suppress inter-user interference, the precoder associated with user $k$ must satisfy
{
\begin{equation}
\tilde{\bm{H}}_k \bm{W}_k
=
\bm{0}.
\label{eq:interference_nulling}
\end{equation}
}

The next section introduces the proposed partitioned Sub-MiLAC architecture for user-specific projection-based MU-MIMO precoding.

\begingroup
\begin{figure}
    \begin{center}
		\makebox[\linewidth][c]{\begin{circuitikz}[
    scale=0.45,
    rotate=90,
    transform shape,
    every node/.style={scale=1.5, rotate=-90},
    circuit ee IEC,
    every path/.style={line width=1pt},
    every resistor/.style={circuit symbol size=width 1.5 height 0.6, line width=1pt, draw=black, fill=white_soft},
    red resistor/.style={resistor={draw=red, fill=red_light}}]

\draw [ color=yellow_mid, draw opacity=0 , fill=yellow_light, fill opacity=1, line width=1pt , rounded corners = 12.0] (13.625,17.875) rectangle (16.125,-0.25);
\draw [ color=blue_light, draw opacity=0 , fill=pink_light, fill opacity=1, line width=1pt , rounded corners = 12.0] (6.125,17.875) rectangle (8.875,-0.25);
\draw [ fill=blue_light, fill opacity=1, line width=1pt , rounded corners = 12.0] (9.25,17.875) rectangle (13.25,-0.25);
\node [font=\small, color=pink_mid, text opacity=1, , inner xsep=0.080cm, inner ysep=0.085cm, rounded corners=0.020cm] at (6.5,8.8125) {Input};
\node [font=\small, color=yellow_mid, text opacity=1, , inner xsep=0.080cm, inner ysep=0.085cm, rounded corners=0.020cm] at (15.65,8.8125) {Output};
\node [font=\normalsize, rotate=90, inner xsep=0.080cm, inner ysep=0.085cm, rounded corners=0.020cm] at (11.25,18.5) {MiLAC};
\draw [ color=white_soft, draw opacity=1 , fill=white_soft, fill opacity=0.68, line width=1pt , rounded corners = 12.0, dashed] (7,17.25) rectangle  (15.25,12.5);
\draw [ color=pink_mid, draw opacity=1, line width=1pt, short] (9.25,16.5) -- (8.5,16.5);
\draw [ color=pink_mid, draw opacity=1, line width=1pt, short] (9.25,15.25) -- (8.5,15.25);
\draw [ color=yellow_mid, draw opacity=1, line width=1pt, short] (14,16.5) -- (13.25,16.5);
\draw [ color=yellow_mid, draw opacity=1, line width=1pt, short] (14,15.25) -- (13.25,15.25);
\draw [ color=yellow_mid, draw opacity=1, line width=1pt, short] (14,13.375) -- (13.25,13.375);
\node [font=\small, color=blue_mid, text opacity=1, , inner xsep=0.080cm, inner ysep=0.085cm, rounded corners=0.020cm] at (11.25,14.875) {SubMiLAC 1};
\node [font=\small, color=pink_mid, text opacity=1, , inner xsep=0.080cm, inner ysep=0.085cm, rounded corners=0.020cm] at (7.75,16.5) {$p_{1,1}$};
\draw [ color=pink_mid, draw opacity=1, line width=1pt, short] (9.25,13.375) -- (8.5,13.375);
\node [font=\small, color=pink_mid, text opacity=1, , inner xsep=0.080cm, inner ysep=0.085cm, rounded corners=0.020cm] at (7.75,15.25) {$p_{1,2}$};
\node [font=\small, color=pink_mid, text opacity=1, , inner xsep=0.080cm, inner ysep=0.085cm, rounded corners=0.020cm] at (7.75,13.375) {$p_{1,N_T}$};
\node [font=\small, color=pink_mid, text opacity=1, , inner xsep=0.080cm, inner ysep=0.085cm, rounded corners=0.020cm] at (7.375,13.375) {};
\node [font=\small, color=yellow_mid, text opacity=1, , inner xsep=0.080cm, inner ysep=0.085cm, rounded corners=0.020cm] at (14.5,16.5) {$p_{1,1}$};
\node [font=\small, color=yellow_mid, text opacity=1, , inner xsep=0.080cm, inner ysep=0.085cm, rounded corners=0.020cm] at (14.5,15.25) {$p_{1,2}$};
\node [font=\small, color=yellow_mid, text opacity=1, , inner xsep=0.080cm, inner ysep=0.085cm, rounded corners=0.020cm] at (14.375,13.375) {};
\node [font=\small, color=yellow_mid, text opacity=1, , inner xsep=0.080cm, inner ysep=0.085cm, rounded corners=0.020cm] at (14.5,13.375) {$p_{1,N_T}$};
\node [font=\small, color=pink_mid, text opacity=1, inner sep=0pt] at (8.5,14.25) {$\dots$};
\node [font=\small, color=yellow_mid, text opacity=1, inner sep=0pt] at (14,14.25) {$\dots$};
\draw [ color=white_soft, draw opacity=1 , fill=white_soft, fill opacity=0.68, line width=1pt , rounded corners = 12.0, dashed] (7,11.125) rectangle  (15.25,6.375);
\draw [ color=pink_mid, draw opacity=1, line width=1pt, short] (9.25,10.375) -- (8.5,10.375);
\draw [ color=pink_mid, draw opacity=1, line width=1pt, short] (9.25,9.125) -- (8.5,9.125);
\draw [ color=yellow_mid, draw opacity=1, line width=1pt, short] (14,10.375) -- (13.25,10.375);
\draw [ color=yellow_mid, draw opacity=1, line width=1pt, short] (14,9.125) -- (13.25,9.125);
\draw [ color=yellow_mid, draw opacity=1, line width=1pt, short] (14,7.25) -- (13.25,7.25);
\node [font=\small, color=blue_mid, text opacity=1, , inner xsep=0.080cm, inner ysep=0.085cm, rounded corners=0.020cm] at (11.25,8.75) {SubMiLAC 2};
\node [font=\small, color=pink_mid, text opacity=1, , inner xsep=0.080cm, inner ysep=0.085cm, rounded corners=0.020cm] at (7.75,10.375) {$p_{2,1}$};
\draw [ color=pink_mid, draw opacity=1, line width=1pt, short] (9.25,7.25) -- (8.5,7.25);
\node [font=\small, color=pink_mid, text opacity=1, , inner xsep=0.080cm, inner ysep=0.085cm, rounded corners=0.020cm] at (7.75,9.125) {$p_{2,2}$};
\node [font=\small, color=pink_mid, text opacity=1, , inner xsep=0.080cm, inner ysep=0.085cm, rounded corners=0.020cm] at (7.75,7.25) {$p_{2,N_T}$};
\node [font=\small, color=pink_mid, text opacity=1, , inner xsep=0.080cm, inner ysep=0.085cm, rounded corners=0.020cm] at (7.375,7.25) {};
\node [font=\small, color=yellow_mid, text opacity=1, , inner xsep=0.080cm, inner ysep=0.085cm, rounded corners=0.020cm] at (14.5,10.375) {$p_{2,1}$};
\node [font=\small, color=yellow_mid, text opacity=1, , inner xsep=0.080cm, inner ysep=0.085cm, rounded corners=0.020cm] at (14.5,9.125) {$p_{2,2}$};
\node [font=\small, color=yellow_mid, text opacity=1, , inner xsep=0.080cm, inner ysep=0.085cm, rounded corners=0.020cm] at (14.375,7.25) {};
\node [font=\small, color=yellow_mid, text opacity=1, , inner xsep=0.080cm, inner ysep=0.085cm, rounded corners=0.020cm] at (14.5,7.25) {$p_{2,N_T}$};
\node [font=\small, color=pink_mid, text opacity=1, inner sep=0pt] at (8.5,8.125) {$\dots$};
\node [font=\small, color=yellow_mid, text opacity=1, inner sep=0pt] at (14,8.125) {$\dots$};
\draw[line width=1pt, color=milacred] (10,12.5) -- (10,11.125);

\draw[line width=1pt, color=milacred] (11,12.5) -- (11,11.125);

\draw[line width=1pt, color=milacred] (12.5,12.5) -- (12.5,11.125);

\node [font=\small, color=milacred, text opacity=1, , inner xsep=0.080cm, inner ysep=0.085cm, rounded corners=0.000cm] at (11.85,11.8125) {$\vdots$};
\draw [ color=white_soft, draw opacity=1 , fill=white_soft, fill opacity=0.68, line width=1pt , rounded corners = 12.0, dashed] (7,5.125) rectangle  (15.25,0.375);
\draw [ color=pink_mid, draw opacity=1, line width=1pt, short] (9.25,4.375) -- (8.5,4.375);
\draw [ color=pink_mid, draw opacity=1, line width=1pt, short] (9.25,3.125) -- (8.5,3.125);
\draw [ color=yellow_mid, draw opacity=1, line width=1pt, short] (14,4.375) -- (13.25,4.375);
\draw [ color=yellow_mid, draw opacity=1, line width=1pt, short] (14,3.125) -- (13.25,3.125);
\draw [ color=yellow_mid, draw opacity=1, line width=1pt, short] (14,1.25) -- (13.25,1.25);
\node [font=\small, color=blue_mid, text opacity=1, , inner xsep=0.080cm, inner ysep=0.085cm, rounded corners=0.020cm] at (11.25,2.75) {SubMiLAC K};
\node [font=\small, color=pink_mid, text opacity=1, , inner xsep=0.080cm, inner ysep=0.085cm, rounded corners=0.020cm] at (7.75,4.375) {$p_{K,1}$};
\draw [ color=pink_mid, draw opacity=1, line width=1pt, short] (9.25,1.25) -- (8.5,1.25);
\node [font=\small, color=pink_mid, text opacity=1, , inner xsep=0.080cm, inner ysep=0.085cm, rounded corners=0.020cm] at (7.75,3.125) {$p_{K,2}$};
\node [font=\small, color=pink_mid, text opacity=1, , inner xsep=0.080cm, inner ysep=0.085cm, rounded corners=0.020cm] at (7.75,1.25) {$p_{K,N_T}$};
\node [font=\small, color=pink_mid, text opacity=1, , inner xsep=0.080cm, inner ysep=0.085cm, rounded corners=0.020cm] at (7.375,1.25) {};
\node [font=\small, color=yellow_mid, text opacity=1, , inner xsep=0.080cm, inner ysep=0.085cm, rounded corners=0.020cm] at (14.5,4.375) {$p_{K,1}$};
\node [font=\small, color=yellow_mid, text opacity=1, , inner xsep=0.080cm, inner ysep=0.085cm, rounded corners=0.020cm] at (14.5,3.125) {$p_{K,2}$};
\node [font=\small, color=yellow_mid, text opacity=1, , inner xsep=0.080cm, inner ysep=0.085cm, rounded corners=0.020cm] at (14.375,1.25) {};
\node [font=\small, color=yellow_mid, text opacity=1, , inner xsep=0.080cm, inner ysep=0.085cm, rounded corners=0.020cm] at (14.5,1.25) {$p_{K,N_T}$};
\node [font=\small, color=pink_mid, text opacity=1, inner sep=0pt] at (8.5,2.125) {$\dots$};
\node [font=\small, color=yellow_mid, text opacity=1, inner sep=0pt] at (14,2.125) {$\dots$};
\node [font=\small, color=blue_mid, text opacity=1, inner sep=0pt] at (11.25,5.75) {$\dots$};
\draw [ color=pink_mid, draw opacity=1 , fill=white_soft, fill opacity=1, line width=1pt ] (8.375,16.5) circle (0.125cm);
\draw [ color=pink_mid, draw opacity=1 , fill=white_soft, fill opacity=1, line width=1pt ] (8.375,15.25) circle (0.125cm);
\draw [ color=pink_mid, draw opacity=1 , fill=white_soft, fill opacity=1, line width=1pt ] (8.375,13.375) circle (0.125cm);
\draw [ color=pink_mid, draw opacity=1 , fill=white_soft, fill opacity=1, line width=1pt ] (8.375,10.375) circle (0.125cm);
\draw [ color=pink_mid, draw opacity=1 , fill=white_soft, fill opacity=1, line width=1pt ] (8.375,9.125) circle (0.125cm);
\draw [ color=pink_mid, draw opacity=1 , fill=white_soft, fill opacity=1, line width=1pt ] (8.375,7.25) circle (0.125cm);
\draw [ color=pink_mid, draw opacity=1 , fill=white_soft, fill opacity=1, line width=1pt ] (8.375,4.375) circle (0.125cm);
\draw [ color=pink_mid, draw opacity=1 , fill=white_soft, fill opacity=1, line width=1pt ] (8.375,3.125) circle (0.125cm);
\draw [ color=pink_mid, draw opacity=1 , fill=white_soft, fill opacity=1, line width=1pt ] (8.375,1.25) circle (0.125cm);
\end{circuitikz}}
    {\footnotesize \\(a) Partitioned MiLAC architecture composed of $K$ independent user-specific Sub-MiLAC subnetworks.}
    \end{center}
    
	\vspace{0cm}
    \begin{center}
		\makebox[\linewidth][c]{\begin{circuitikz}[
    scale=0.45,
    every node/.style={scale=0.7},
    circuit ee IEC,
    every path/.style={line width=1pt},
    every resistor/.style={circuit symbol size=width 1.5 height 0.6, line width=1pt, draw=black, fill=white_soft},
    red resistor/.style={resistor={draw=red, fill=red_light}}]
\newcommand{\submilacYlabel}[4]{%
{\small
$
y_{
{\color{#1}k,#2};\,
{\color{#3}k,#4}
}
$
}
}
\draw [ color=white_soft, draw opacity=0 , fill=yellow_light, fill opacity=1, rounded corners = 12.0] (32.625,20.25) rectangle (35.5,10.875);
\draw [ color=white_soft, draw opacity=0 , fill=pink_light, fill opacity=1, rounded corners = 12.0] (15.25,20.25) rectangle (18.125,10.875);
\draw [ color=black, draw opacity=0 , fill=blue_light, fill opacity=1] (18.75,20.25) rectangle (32,10.875);
\draw [line width=1pt, short] (32,20.25) -- (32,10.875);
\draw [line width=1pt, short] (18.75,20.25) -- (18.75,10.875);
\draw [ color=white_soft, draw opacity=1 , fill=white_soft, fill opacity=0.58, line width=1pt , rounded corners = 12.0, dashed] (16.25,12.125) rectangle  (34.5,10.25);
\draw [ color=white_soft, draw opacity=1 , fill=white_soft, fill opacity=0.58, line width=1pt , rounded corners = 12.0, dashed] (16.25,20.875) rectangle  (34.5,19);
\draw [ color=white_soft, draw opacity=1 , fill=white_soft, fill opacity=0.58, line width=1pt , rounded corners = 12.0, dashed] (16.25,18.375) rectangle  (34.5,12.75);
\draw [line width=1pt, short] (17.5,17.125) -- (25.25,17.125);
\draw [line width=1pt, short] (33.25,17.125) -- (28,17.125);
\draw [line width=1pt, short] (17.5,15.25) -- (25.125,15.25);
\draw [line width=1pt, short] (25.125,15.25) arc[start angle=180, end angle=0, radius=0.1261];
\draw [line width=1pt, short] (25.3772,15.25) -- (33.25,15.25);
\draw [line width=1pt, short] (17.5,13.375) -- (23.625,13.375);
\draw [line width=1pt, short] (33.25,13.375) -- (28,13.375);
\node[font=\small, color=milacred, text opacity=1] at (28.55,16.1875) {$\boldsymbol{\cdots}$};
\node[font=\small, color=milacred, text opacity=1] at (29.5,16.1875) {$\boldsymbol{\cdots}$};
\node[font=\small, color=milacred, text opacity=1] at (30.5,16.1875) {$\boldsymbol{\cdots}$};
\draw [line width=1pt, short] (23.625,13.375) -- (28.125,13.375);
\draw [line width=1pt, short] (25.25,17.125) -- (28,17.125);
\draw[line width=1pt] (22.5,17.125) to[resistor] node[midway, right, yshift=-1pt, fill=white_soft, fill opacity=0, text opacity=1, inner sep=3.5pt, font=\small] {\submilacYlabel{pink_mid}{1}{pink_mid}{2}} (22.5,15.25);
\draw[line width=1pt] (22.5,15.25) to[resistor] node[midway, right, yshift=-1pt, fill=white_soft, fill opacity=0, text opacity=1, inner sep=3.5pt, font=\small] {\submilacYlabel{pink_mid}{2}{pink_mid}{3}} (22.5,13.375);
\draw[line width=1pt]
(25.25,17.125) -- (25.25,16.6);

\draw[line width=1pt]
(25.25,16.6)
to[resistor]
node[midway,right,
     yshift=-1pt,
     fill=white_soft,
     fill opacity=0,
     text opacity=1,
     inner sep=3.5pt,
     font=\small]
{\submilacYlabel{pink_mid}{1}{pink_mid}{3}}
(25.25,15.9);

\draw[line width=1pt]
(25.25,15.9) -- (25.25,13.375);

\draw [ color=milacred, draw opacity=1, line width=1pt, short] (31,10.875) -- (31,13.25);
\draw [ color=milacred, draw opacity=1, line width=1pt, short] (31,13.5) arc[start angle=97.74, end angle=-97.74, radius=0.1261];
\draw [ color=milacred, draw opacity=1, line width=1pt, short] (31,17.25) arc[start angle=97.74, end angle=-97.74, radius=0.1261];
\draw [ color=milacred, draw opacity=1, line width=1pt, short] (31,17.25) -- (31,20.25);
\draw [ color=milacred, draw opacity=1, line width=1pt, short] (30,10.875) -- (30,15.125);
\draw [ color=milacred, draw opacity=1, line width=1pt, short] (30,15.125) arc[start angle=-90, end angle=90, radius=0.1261];
\draw [ color=milacred, draw opacity=1, line width=1pt, short] (30,15.375) -- (30,17);
\draw [ color=milacred, draw opacity=1, line width=1pt, short] (30,17) arc[start angle=-90, end angle=90, radius=0.1261];
\draw [ color=milacred, draw opacity=1, line width=1pt, short] (30,17.25) -- (30,20.25);
\draw [ color=milacred, draw opacity=1, line width=1pt, short] (31,13.5) -- (31,15.125);
\draw [ color=milacred, draw opacity=1, line width=1pt, short] (31,15.125) arc[start angle=-90, end angle=90, radius=0.1261];
\draw [ color=milacred, draw opacity=1, line width=1pt, short] (31,15.375) -- (31,17);
\draw[line width=1pt]
(19.75,17.125)
to[resistor]
node[midway, right, yshift=-1pt, fill=white_soft, fill opacity=0, text opacity=1, inner sep=3.5pt, font=\small] {\submilacYlabel{pink_mid}{1}{pink_mid}{1}}
(19.75,18.45);

\draw[-{Triangle[open,scale=1,fill=white_soft]}, line width=1pt]
(19.75,18.45) -- (19.75,18.625);

\draw[line width=1pt]
(19.75,13.375)
to[resistor]
node[midway, right, yshift=-1pt, fill=white_soft, fill opacity=0, text opacity=1, inner sep=3.5pt, font=\small] {\submilacYlabel{pink_mid}{3}{pink_mid}{3}}
(19.75,14.70);

\draw[-{Triangle[open,scale=1,fill=white_soft]}, line width=1pt]
(19.75,14.70) -- (19.75,14.875);

\draw[line width=1pt]
(19.75,15.25)
to[resistor]
node[midway, right, yshift=-1pt, fill=white_soft, fill opacity=0, text opacity=1, inner sep=3.5pt, font=\small] {\submilacYlabel{pink_mid}{2}{pink_mid}{2}}
(19.75,16.575);

\draw[-{Triangle[open,scale=1,fill=white_soft]}, line width=1pt]
(19.75,16.575) -- (19.75,16.75);

\node [font=\small, color=blue_mid, text opacity=1, , inner xsep=0.080cm, inner ysep=0.085cm, rounded corners=0.020cm] at (25.375,17.75) {SubMiLAC k};
\draw [ color=milacred, draw opacity=1, line width=1pt, short] (28,10.875) -- (28,13.25);
\draw [ color=milacred, draw opacity=1, line width=1pt, short] (28,13.25) arc[start angle=-90, end angle=90, radius=0.1261];
\draw [ color=milacred, draw opacity=1, line width=1pt, short] (28,13.5) -- (28,15.125);
\draw [ color=milacred, draw opacity=1, line width=1pt, short] (28,15.125) arc[start angle=-90, end angle=90, radius=0.1261];
\draw [ color=milacred, draw opacity=1, line width=1pt, short] (28,15.375) -- (28,20.25);
\draw [ color=milacred, draw opacity=1, line width=1pt, short] (29,10.875) -- (29,13.25);
\draw [ color=milacred, draw opacity=1, line width=1pt, short] (29,13.25) arc[start angle=-90, end angle=90, radius=0.1261];
\draw [ color=milacred, draw opacity=1, line width=1pt, short] (29,13.5) -- (29,17);
\draw [ color=milacred, draw opacity=1, line width=1pt, short] (29,17) arc[start angle=-90, end angle=90, radius=0.1261];
\draw [ color=milacred, draw opacity=1, line width=1pt, short] (29,17.25) -- (29,20.25);
\node [font=\small, color=blue_mid, text opacity=1, , inner xsep=0.080cm, inner ysep=0.085cm, rounded corners=0.020cm] at (25.375,19.5) {SubMiLAC $k-1$};
\node [font=\small, color=blue_mid, text opacity=1, , inner xsep=0.080cm, inner ysep=0.085cm, rounded corners=0.020cm] at (25.375,11.5) {SubMiLAC $k+1$};
\node [font=\small, color=yellow_mid, text opacity=1, , inner xsep=0.080cm, inner ysep=0.085cm, rounded corners=0.020cm] at (33.625,17.625) {$p_{k,1}$};
\node [font=\small, color=yellow_mid, text opacity=1, , inner xsep=0.080cm, inner ysep=0.085cm, rounded corners=0.020cm] at (33.625,15.75) {$p_{k,2}$};
\node [font=\small, color=yellow_mid, text opacity=1, , inner xsep=0.080cm, inner ysep=0.085cm, rounded corners=0.020cm] at (33.625,13.875) {$p_{k,3}$};
\node [font=\small, color=pink_mid, text opacity=1, , inner xsep=0.080cm, inner ysep=0.085cm, rounded corners=0.020cm] at (17.125,17.625) {$p_{k,1}$};
\node [font=\small, color=pink_mid, text opacity=1, , inner xsep=0.080cm, inner ysep=0.085cm, rounded corners=0.020cm] at (17.125,15.75) {$p_{k,2}$};
\node [font=\small, color=pink_mid, text opacity=1, , inner xsep=0.080cm, inner ysep=0.085cm, rounded corners=0.020cm] at (17.125,13.875) {$p_{k,3}$};
\node [font=\small, color=pink_mid, text opacity=1, , rotate around={90:(0,0)}, inner xsep=0.080cm, inner ysep=0.085cm, rounded corners=0.020cm] at (15.75,15.625) {Input};
\node [font=\small, color=yellow_mid, text opacity=1, , rotate around={-90:(0,0)}, inner xsep=0.080cm, inner ysep=0.085cm, rounded corners=0.020cm] at (35,15.5) {Output};
\draw [ fill=white_soft, fill opacity=1, line width=1pt ] (25.25,17.125) circle (0.125cm);
\draw [ fill=white_soft, fill opacity=1, line width=1pt ] (25.25,13.375) circle (0.125cm);
\draw [ fill=white_soft, fill opacity=1, line width=1pt ] (22.5,17.125) circle (0.125cm);
\draw [ fill=white_soft, fill opacity=1, line width=1pt ] (22.5,15.25) circle (0.125cm);
\draw [ fill=white_soft, fill opacity=1, line width=1pt ] (22.5,13.375) circle (0.125cm);
\draw [ fill=white_soft, fill opacity=1, line width=1pt ] (17.5,17.125) circle (0.125cm);
\draw [ fill=white_soft, fill opacity=1, line width=1pt ] (17.5,15.25) circle (0.125cm);
\draw [ fill=white_soft, fill opacity=1, line width=1pt ] (17.5,13.375) circle (0.125cm);
\draw [ fill=white_soft, fill opacity=1, line width=1pt ] (29,15.25) circle (0.125cm);
\draw [ fill=white_soft, fill opacity=1, line width=1pt ] (30,13.375) circle (0.125cm);
\draw [ fill=white_soft, fill opacity=1, line width=1pt ] (19.75,13.375) circle (0.125cm);
\draw [ fill=white_soft, fill opacity=1, line width=1pt ] (19.75,15.25) circle (0.125cm);
\draw [ fill=white_soft, fill opacity=1, line width=1pt ] (19.75,17.125) circle (0.125cm);
\draw [ fill=white_soft, fill opacity=1, line width=1pt ] (28,17.125) circle (0.125cm);
\end{circuitikz}}
    {\footnotesize \\(b) Internal structure and local port partition of a single Sub-MiLAC subnetwork.}
    \end{center}
    
    \vspace{-0.3cm}
    \caption{\label{fig:submilac_architecture}
Proposed partitioned MiLAC architecture for MU-MIMO precoding.
}
\vspace{-0.2cm}
\end{figure}
\endgroup
\section{Proposed Method}

\begingroup
Since a single global MiLAC transformation couples the contributions associated with different users, the transmitter-side MiLAC is partitioned into $K$ independent user-specific subnetworks to enable user-specific interference-aware processing. Each Sub-MiLAC operates in a full-port excitation mode: its $N_T$ physical ports are simultaneously excited and observed, with the excitation represented by incident waves and the output by the resulting port voltages \cite{nerini2025part1,nerini2026hybrid}. At port $n$, the voltage satisfies $x_n=a_n+b_n$, where $a_n$ and $b_n$ are the incident- and reflected-wave amplitudes, respectively. Thus, the complete architecture contains $P=KN_T$ physical ports.

Figure~\ref{fig:submilac_architecture} illustrates the proposed architecture. The input and output terminals drawn on opposite sides of each Sub-MiLAC are included solely to clarify the signal flow. As shown in Fig.~\ref{fig:submilac_architecture}(b), each right-side terminal is short-circuited to its corresponding left-side terminal, so every pair represents a single physical port rather than distinct input and output terminals. A port is denoted by $p_{m,i}$, where $m\in\{1,\ldots,K\}$ identifies the subnetwork (or user) and $i\in\{1,\ldots,N_T\}$ identifies the local physical port within that subnetwork.

Ideally, the subnetworks are electromagnetically isolated by enforcing zero
inter-subnetwork admittances, represented by the red interconnections in
Fig.~\ref{fig:submilac_architecture}. In practice, this condition is
approximated by ensuring that the mutual admittances between ports belonging
to different subnetworks are sufficiently small, i.e.,
\begin{equation}
Y_{m,i;\,n,j}
\approx 0,
\qquad
m \neq n,
\label{eq:inter_subnetwork_admittance}
\end{equation}
where $Y_{m,i;\,n,j}$ denotes the mutual admittance between port $i$ of
subnetwork $m$ and port $j$ of subnetwork $n$. The subnetwork indices
precede the corresponding local port indices.

Each Sub-MiLAC is characterized by a self-transformation matrix
$\bm{P}_k\in\mathbb{C}^{N_T\times N_T}$, whereas the electromagnetic interaction between distinct
subnetworks is described by the coupling matrices
$\bm{C}_{m,n}\in\mathbb{C}^{N_T\times N_T}$.
Under the weak-coupling assumption, $\bm{C}_{m,n}\approx\mathbf0$ for
$m\neq n$, the global MiLAC transformation becomes approximately block
diagonal.

Accordingly, the complete partitioned MiLAC system can be represented as
\begin{equation}
\begin{bmatrix}
\bm{P}_1 & \bm{C}_{1,2} & \cdots & \bm{C}_{1,K} \\
\bm{C}_{2,1} & \bm{P}_2 & \cdots & \bm{C}_{2,K} \\
\vdots & \vdots & \ddots & \vdots \\
\bm{C}_{K,1} & \bm{C}_{K,2} & \cdots & \bm{P}_K
\end{bmatrix}
\begin{bmatrix}
\bm{x}_1 \\
\bm{x}_2 \\
\vdots \\
\bm{x}_K
\end{bmatrix}
=
\begin{bmatrix}
\bm{u}_1 \\
\bm{u}_2 \\
\vdots \\
\bm{u}_K
\end{bmatrix}.
\end{equation}

For the $k$-th Sub-MiLAC, the transmitted signal contribution is given by
\begin{equation}
\bm{x}_k=\bm{P}_k^{-1}\bm{u}_k,
\label{eq:milac_io}
\end{equation}

Equation~\eqref{eq:milac_io} establishes the fundamental input--output
relationship of each Sub-MiLAC, showing that any desired invertible linear
transformation can be realized by appropriately selecting the transformation
matrix $\bm{P}_k$. Consequently, the transmitted contribution
$\bm{x}_k$ is obtained by applying the inverse transformation
$\bm{P}_k^{-1}$ to the excitation vector $\bm{u}_k$.

Moreover, the proposed architecture naturally supports adaptive hardware partitioning. Since the total number of MiLAC ports defines the available hardware resources, the same physical MiLAC can be dynamically partitioned into different Sub-MiLAC configurations by reconfiguring the interconnections among ports. Specifically, setting the admittances between selected subnetworks to approximately zero effectively decouples them, allowing the number and dimensions of the resulting Sub-MiLACs to be adapted to the number of users, the number of transmitted streams, hardware resource allocation, and interference-mitigation requirements.

The realization of a practical mechanism capable of dynamically adjusting these
interconnection admittances while preserving the desired network response is,
however, beyond the scope of this work and constitutes an interesting direction
for future research.

\subsection{Dimension Matching}
\label{subsec:dimension_matching}
Since the excitation vector has a fixed dimension imposed by the all-input mode of the proposed MiLAC architecture, its dimension may not match that of the symbol vector. Consequently, the excitation vector must be mapped to the appropriate dimension required by the output vector of each subnetwork.

This preprocessing is performed through a reduced Hermitian channel matrix
$\bar{\bm{H}}_k^{H}\in\mathbb{C}^{N_T\times D_k}$, obtained by selecting the
columns of $\bm{H}_k^{H}$ associated with the transmitted streams. The resulting
MiLAC excitation vector is given by
\begin{equation}
\bm{u}_k=\bar{\bm{H}}_k^{H}\bm{s}_k,
\label{eq:input_embedding}
\end{equation}
which satisfies $\bm{u}_k\in\mathbb{C}^{N_T\times1}$ and therefore matches the
dimensionality required. Once this
dimension-compatible excitation has been formed, the remaining task is to
configure $\bm{P}_k$ so that the corresponding response $\bm{x}_k$ satisfies
the user-specific interference-nulling constraint.

\subsection{Projection-Based Reformulation}

To connect the Sub-MiLAC relation in~\eqref{eq:milac_io} with the MU-MIMO
constraint in~\eqref{eq:interference_nulling}, we employ the orthogonal
projector onto the null space of $\tilde{\bm{H}}_k$, thereby avoiding explicit
null-space basis computation. Assuming that $\tilde{\bm{H}}_k$ has full row
rank, this projector is given by
\begin{equation}
\bm{\Pi}_k
=
\bm{I}_{N_T}
-
\tilde{\bm{H}}_k^{H}
\left(
\tilde{\bm{H}}_k
\tilde{\bm{H}}_k^{H}
\right)^{-1}
\tilde{\bm{H}}_k,
\label{eq:null_projector}
\end{equation}
which satisfies $\tilde{\bm{H}}_k\bm{\Pi}_k=\bm{0}$. Consequently, any vector
projected by $\bm{\Pi}_k$ belongs to the null space of the complementary user
channel, thereby eliminating inter-user interference.

However, since $\bm{\Pi}_k$ is an orthogonal projector, it is idempotent and
rank deficient, and therefore singular. As a result, it cannot be directly
employed in the proposed MiLAC synthesis, which requires an invertible
transformation matrix.

To overcome this limitation, a regularized projection operator is introduced as
\begin{equation}
\bm{\Pi}_{k,\epsilon}
=
\bm{\Pi}_k
+
\epsilon\bm{I}_{N_T},
\qquad
\epsilon>0,
\label{eq:regularized_projector}
\end{equation}
where $\epsilon$ is a sufficiently small regularization parameter. The
transformation matrix synthesized by the $k$-th Sub-MiLAC is then chosen as
\begin{equation}
\bm{P}_k
=
\bm{\Pi}_{k,\epsilon}^{-1},
\label{eq:Pk_regularized}
\end{equation}
such that the analog network realizes the desired regularized projection
operation.
\endgroup

\section{Simulation Results}
\begin{figure}[hbt]
\centering
\includegraphics[
    width=0.48\textwidth,
    trim={0cm 0cm 0cm 0.38cm},
    clip
]{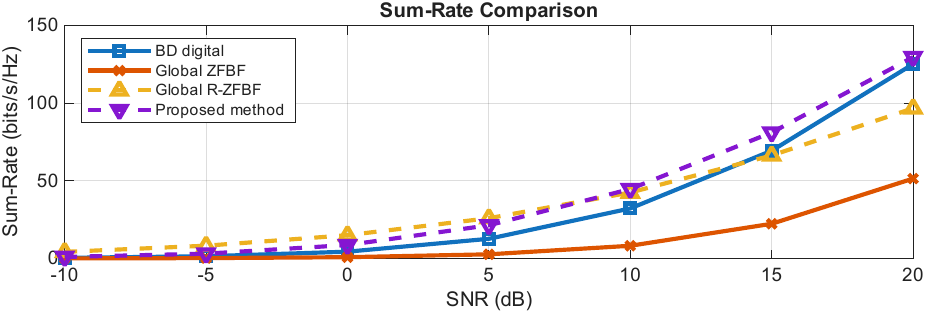}
\vspace{-0.15cm}

{\footnotesize (a) Average achievable sum-rate versus SNR.}
\vspace{0.3cm}

\includegraphics[
    width=0.48\textwidth,
    trim={0cm 0cm 0cm 0.38cm},
    clip
]{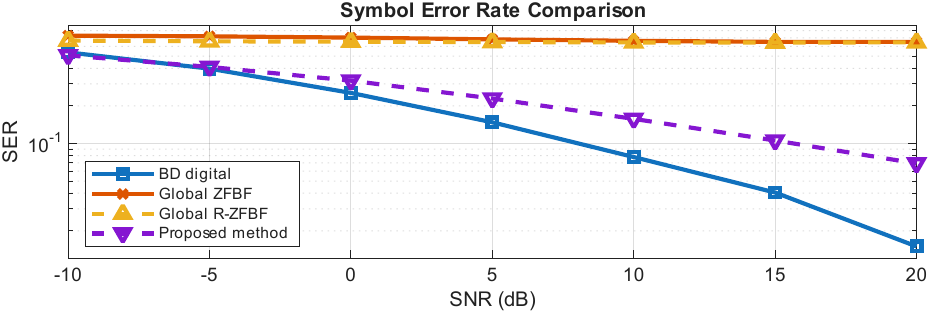}
\vspace{-0.15cm}

{\footnotesize (b) Symbol Error Rate (SER) versus SNR.}
\vspace{0.3cm}

\includegraphics[
    width=0.48\textwidth,
    trim={0cm 0cm 0cm 0.38cm},
    clip
]{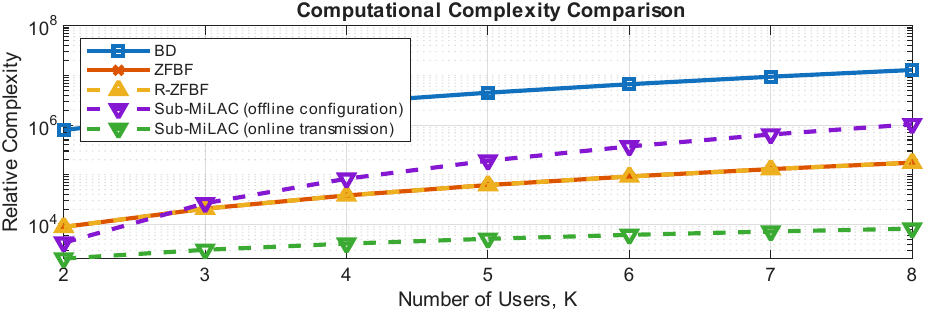}
\vspace{-0.15cm}

{\footnotesize (c) Computational complexity versus number of users.}
\vspace{-0.2cm}
\caption{\label{fig:performance_results}
Performance comparison of the evaluated precoding strategies: (a) achievable sum-rate, (b) SER performance, and (c) computational complexity scaling.
}
\vspace{-0.4cm}
\end{figure}
\subsection{Simulation Setup}

The proposed projection-based Sub-MiLAC precoder is evaluated in a MU-MIMO downlink scenario with \(N_T=64\) transmit antennas and \(K=16\) users. Each user is equipped with \chg{\(N_{R,k}=4\)} receive antennas and is assigned \chg{\(D_k=4\)} transmitted streams, resulting in a total of \(D_{\mathrm{tot}}=64\) transmitted streams.

In this evaluation, ideal Sub-MiLAC decoupling is assumed, such that the inter-subnetwork coupling matrices satisfy $\bm{C}_{m,n}=\bm{0}$ for all $m\neq n$. This assumption allows each Sub-MiLAC to operate independently, so that the user-specific transformations can be implemented in parallel without mutual electromagnetic interference. In practical implementations, residual coupling may arise from finite isolation and parasitic effects; its impact, as well as coupling-aware network design and reconfigurable decoupling mechanisms, is left for future investigation.

The wireless channel follows a geometric MIMO model with randomly generated propagation paths. For each SNR point, the results are averaged over \(1000\) independent Monte Carlo channel realizations.

The proposed method is compared with fully digital BD, fully digital ZFBF, and R-ZFBF.

\subsection{Achievable Sum-Rate and Symbol Error Rate Performance}

The achievable sum-rate is computed as
{
\[
R_{\mathrm{sum}}
=
\sum_{k=1}^{K}
\log_2\det(
\mathbf{I}_{d_k}
+
(\mathbf{R}_{k,n})^{-1}\mathbf{S}_k
),
\]
Here, $\mathbf{S}_k$ is the desired-signal covariance matrix, while $\mathbf{R}_{k,n}$ is the interference-plus-noise covariance matrix for user $k$; the second subscript $n$ identifies the interference-plus-noise term.
}

Fig.~\ref{fig:performance_results}(a) presents the achievable sum-rate performance of the evaluated precoding strategies. The proposed Sub-MiLAC precoder achieves spectral efficiency close to fully-digital BD across the evaluated SNR range. At low SNR, the matched-filter component improves desired-signal enhancement, while at high SNR the absence of explicit stream orthogonalization introduces a small performance gap relative to BD.

Fig.~\ref{fig:performance_results}(b) shows the corresponding symbol error rate (SER) performance obtained using normalized QPSK signaling and local MMSE detection at each user \cite{tse2005fundamentals}. The SER is computed as
\[
\mathrm{SER}
=
\frac{1}{d}
\sum_{i=1}^{d}
\mathbb{I}(\hat{s}_i \neq s_i),
\]
where $\hat{s}_i$ denotes the detected symbol associated with stream $i$, and $\mathbb{I}(\cdot)$ is the indicator function.

The proposed Sub-MiLAC architecture achieves SER performance close to fully-digital BD, indicating that the analog projection mechanism effectively preserves the user-level interference-nulling structure required for reliable symbol recovery. This behavior is consistent with the interference-suppression properties traditionally associated with block-diagonalization precoding \cite{bd}.

In contrast, globally designed beamforming strategies such as ZFBF and R-ZFBF exhibit noticeably higher SER values, particularly in the moderate- and high-SNR regimes. Although these methods may still achieve competitive spectral efficiency, the results reveal that residual interference leakage significantly degrades symbol detection performance. This behavior reflects the tradeoff between interference suppression and channel conditioning commonly observed in global beamforming formulations \cite{yoo2006optimality}.

Overall, the results demonstrate that the proposed Sub-MiLAC architecture preserves both high spectral efficiency and reliable symbol recovery while remaining compatible with parallel analog implementations and avoiding the computational burden of repeated digital matrix decompositions.

\subsection{Computational Complexity}

Fig.~\ref{fig:performance_results}(c) compares the relative computational complexity of the evaluated precoding strategies as the number of users increases. The comparison focuses on the channel-dependent processing required to obtain the precoding transformations, while signal-dependent operations are considered separately.

For conventional BD, the dominant cost arises from the repeated user-specific SVDs required to obtain the interference-free transmit subspaces. For the $k$-th user, the complementary channel has dimensions approximately $(K-1)N_R\times N_T$. The corresponding SVD-related complexity is approximately $\mathcal{O}(K^3N_R^2N_T)+\mathcal{O}(K^2N_RN_T^2)$, with additional lower-order terms associated with subsequent per-user processing.

Global ZFBF and R-ZFBF operate on the aggregate channel and require the construction and inversion of a $KN_R\times KN_R$ Gram matrix. Their dominant complexity therefore scales as $\mathcal{O}(K^2N_R^2N_T)+\mathcal{O}(K^3N_R^3)$. Regularization in R-ZFBF does not change this asymptotic order.

For the proposed Sub-MiLAC architecture, the offline configuration requires one user-specific projection for each Sub-MiLAC. For comparable $N_R$, forming and inverting the corresponding Gram matrices scales approximately as $\mathcal{O}(K^3N_R^2N_T)+\mathcal{O}(K^4N_R^3)$. This channel-dependent cost is incurred offline and can be reused over multiple transmission intervals.

During online transmission, the digital processing is limited primarily to forming the dimension-matched excitation vectors, with complexity $\mathcal{O}(N_T\sum_{k=1}^{K}D_k)$, which becomes $\mathcal{O}(KN_RN_T)$ for $D_k=N_R$. The subsequent projection is performed by the configured analog subnetworks, eliminating the need to construct and apply a new digital MU-MIMO precoder during transmission. Hence, the main advantage of the proposed architecture is the substantially reduced online digital processing, at the cost of an offline channel-dependent configuration stage.

\section{Conclusion}

The results demonstrate that, under ideal conditions, the proposed Sub-MiLAC architecture enables MU-MIMO transmission while preserving the user-level interference-nulling behavior required by block diagonalization. By decomposing the global analog network into independent user-specific subnetworks, the proposed formulation enables parallel analog processing and replaces explicit null-space basis extraction with projection-based transformations that can be configured into the MiLAC hardware. Although the channel-dependent configuration requires non-negligible offline computation, the resulting analog transformations can be reused over multiple transmission intervals, substantially reducing the online digital processing burden. The proposed architecture achieves spectral efficiency close to fully digital BD while improving symbol reconstruction performance compared with the considered global ZFBF and R-ZFBF approaches. These results highlight Sub-MiLAC architectures as a promising path toward scalable, low-latency, and energy-efficient analog-assisted beamforming for high-dimensional MU-MIMO systems.

{Future work will investigate the practical feasibility of the proposed architecture beyond the ideal assumptions considered here. In particular, it will quantify the effects of residual inter-Sub-MiLAC coupling, imperfect channel estimation, calibration errors, and other analog hardware nonidealities. Further studies will also provide execution-time and implementation-oriented complexity measurements and extend the performance comparisons to recent hybrid-beamforming and microwave analog-computing methods.}

\end{document}